\documentclass[12pt]{article}

\pdfoutput = 1

\usepackage{graphicx}
\usepackage{subcaption}
\usepackage{amstext}
\usepackage{lineno}
\usepackage{url}

\newcommand\pubnumber{ATLAS}
\newcommand\pubdate{January 15, 2024}

\newcommand\blfootnote[1]{%
  \begingroup
  \renewcommand\thefootnote{}\footnote{#1}%
  \addtocounter{footnote}{-1}%
  \endgroup
}

\def\institute{INFN Sezione di Trieste, Gruppo Collegato di Udine\\
Universit\`a degli Studi di Trieste}
\def\authemail{\footnote{Contact: laura.pintucci@cern.ch}}

\def\Title#1{\begin{center} {\Large #1 } \end{center}}
\def\Author#1{\begin{center}{ \sc #1} \end{center}}
\def\Address#1{\begin{center}{ \it #1} \end{center}}

\newcommand\pubblock{\rightline{\begin{tabular}{l} \pubnumber\\
         \pubdate  \end{tabular}}}
\newenvironment{Abstract}{\begin{quotation}  }{\end{quotation}}
\newenvironment{Presented}{\begin{quotation} \begin{center} 
             PRESENTED AT\end{center}\bigskip 
      \begin{center}\begin{large}}{\end{large}\end{center} \end{quotation}}

\def\beq{\begin{equation}}
\def\eeq#1{\label{#1}\end{equation}}
\def\eeqn{\end{equation}}

\def\beqa{\begin{eqnarray}}
\def\eeqa#1{\label{#1}\end{eqnarray}}
\def\eeqan{\end{eqnarray}}

\let\bar=\overbar

\def\Dslash{\not{\hbox{\kern-4pt $D$}}}
\def\dslash{\not{\hbox{\kern-2pt $\del$}}}

\def\msb{{\bar{\ssstyle M \kern -1pt S}}}

\begin{document}
\begin{titlepage}
\pubblock

\vfill
\Title{Measurements of single top quark production processes with the ATLAS and CMS experiments}
\vfill
\Author{ Laura Pintucci\authemail \  on behalf of the ATLAS and CMS Collaborations}
\Address{\institute}
\vfill
\begin{Abstract}
This report contains a brief summary of the latest single top quark production cross-section measurements performed by the ATLAS and CMS collaborations on $pp$ collisions collected during Run 2 of the LHC. Various results for $t-$channel, $s-$channel and $W$ associated single top production are discussed. Particular attention is given to the main techniques used by the analyses and the main systematic uncertainties limiting these precise measurements.

\blfootnote{Copyright 2023 CERN for the benefit of the ATLAS and CMS Collaborations. \\Reproduction of this article or parts of it is allowed as specified in the CC-BY-4.0 license.}
\end{Abstract}
\vfill

\begin{Presented}
$16^\mathrm{th}$ International Workshop on Top Quark Physics\\
(Top2023), 24--29 September, 2023
\end{Presented}
\vfill
\end{titlepage}
\def\thefootnote{\fnsymbol{footnote}}
\setcounter{footnote}{0}
%


\section{Introduction}
\begin{figure}[htpb]
    \centering
    \begin{minipage}{0.34\textwidth}  
        \centering
        \includegraphics[width=0.7\textwidth]{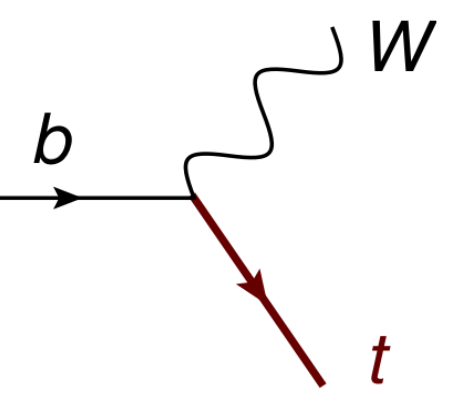}
        \caption{Feynman diagram of a vertex between a $W$ boson, a top quark and a bottom quark.}
        \label{fig:Wtb}
    \end{minipage}
    \hfill
    \begin{minipage}{0.6\textwidth}  
        One of the interesting properties of single top production is that its cross-section is proportional to the square root of the Cabibbo Kobayashi Maskawa (CKM) matrix element $V_{tb}$. Moreover, single top cross-section measurements and charge asymmetry measurements of top quark versus top anti-quark can have an important impact on constraining PDFs, and the three channels are complementary to each other and to $t\bar{t}$ production.
    \end{minipage}
\end{figure}

The top quark can be produced at hadron collider, such as the LHC, via the strong interaction or the EW interaction. In the first case, a top quark and an anti-quark are produced in pair ($t\bar{t}$), while in the second case, a top quark (or anti-quark) is produced, referred to as single top production. 
In the Standard Model (SM) the single top production is a charged current electroweak process, that has a vertex involving a top quark, a bottom quark and a $W$ boson ($Wtb$ vertex), as shown in Figure \ref{fig:Wtb}. Based on how the $Wtb$ vertex is oriented one can have different modes for the single top production: the $t-$channel is the one with the largest cross-section at the LHC and it is characterized by the exchange of a virtual $W$ boson between a light quark and a top quark; the $s-$channel has the smallest cross-section and it is the production and decay of an off-shell $W$ boson; the $W-$associated production has in its final state a top quark and a $W$ boson. 

In the following sections, some of the latest single top quark production cross-section measurements performed by the ATLAS and CMS collaborations on $pp$ collisions collected during Run 2 of the LHC are presented.

\section{Single top $t-$channel production}
In this section, four different $t-$channel measurements are described, two of which are performed using data collected by the CMS experiment~\cite{CMS}, and the other two have been released in the last year using data collected by the ATLAS experiment~\cite{ATLAS}. 

The latest result for the $t-$channel cross-section measurement from the CMS Collaboration uses data at the centre-of-mass energy of $\sqrt{s} = 13$ TeV with a luminosity of $35.9$ fb$^{-1}$~\cite{CMS_1}. In this analysis, events are selected if they contain one electron or muon, two or three jets of which at maximum two of them must be tagged as $b-$jets ($b-$tagged jets). Boosted Decision Trees (BDTs) are used to separate the signal from the background. 
A profile likelihood fit is performed to extract yields of the single top quark, and anti-quark production using the BDTs distributions and the transverse mass of the reconstructed $W$ boson ($m_T(W)$). 
Differential cross-section measurements are extracted with an unfolding at parton and particle level, 
and good agreement is found between data distributions and Monte Carlo (MC) prediction with the 4 and 5 flavour schemes.

\renewcommand{\thefootnote}{\arabic{footnote}}

\begin{figure}[htpb]
    \centering
    \begin{minipage}{0.47\textwidth}
        \centering
        \includegraphics[width=\textwidth]{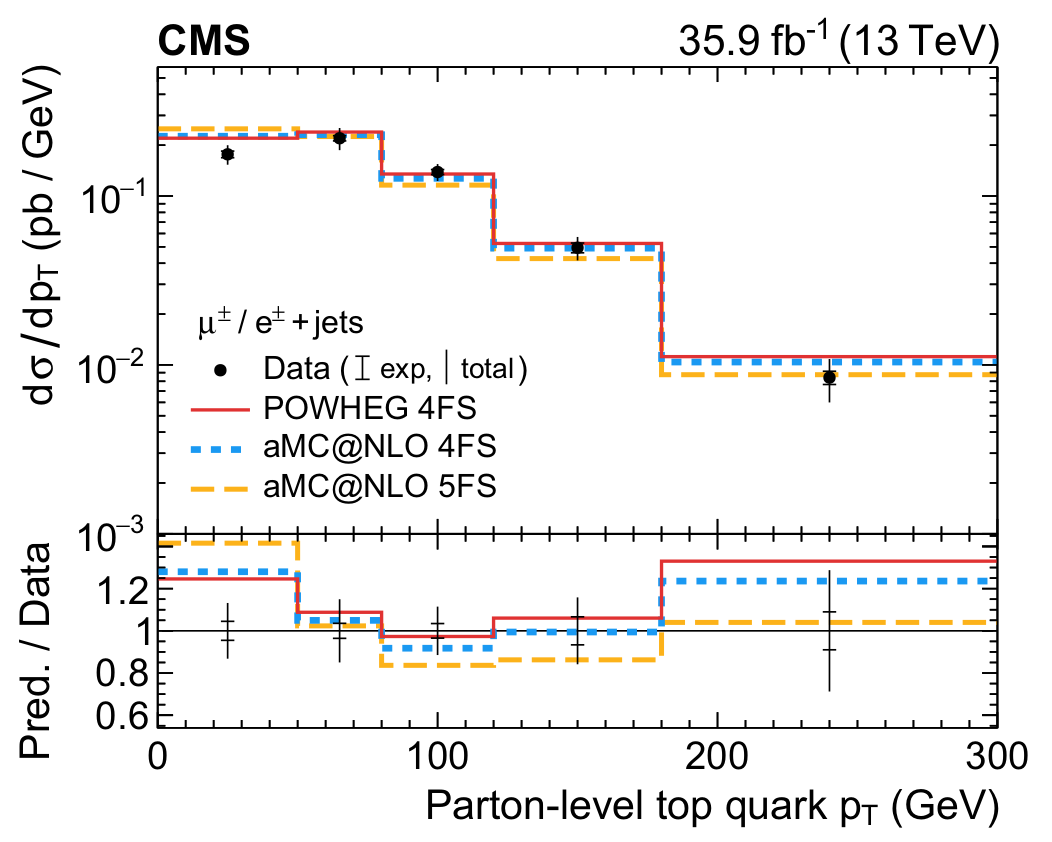}
        \caption{Differential cross section for the $t-$channel at parton level of the top quark $p_T$ extracted by the CMS Collaboration at $\sqrt{s}=13$~TeV~\cite{CMS_1}. 
        }
        \label{fig:CMS_1}
    \end{minipage}
    \hfill
    \begin{minipage}{0.48\textwidth}
        Only for the transverse momentum of the reconstructed top quark and $W$ boson the 5 flavour MC predictions do not agree as well with the data, as shown in Figure \ref{fig:CMS_1}. Differential ratios of the top quark production rates to the sum of the top quark and antiquark rates are compared to predictions from three different PDF sets which are all found to be in good agreement. The spin asymmetry\footnotemark[1] is extracted from the differential cross-section distribution with respect to the polarizing angle at parton level, and it is found to be $0.440 \pm 0.070$ in good agreement with the SM prediction.
    \end{minipage}
\end{figure}

\begin{figure}[htpb]
    \centering
    \begin{subfigure}{0.4\textwidth}
        \includegraphics[width=\textwidth]{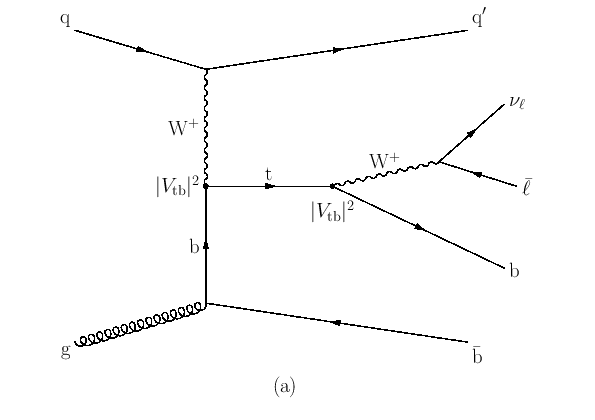} 
    \end{subfigure}
    \begin{subfigure}{0.4\textwidth}
        \includegraphics[width=\textwidth]{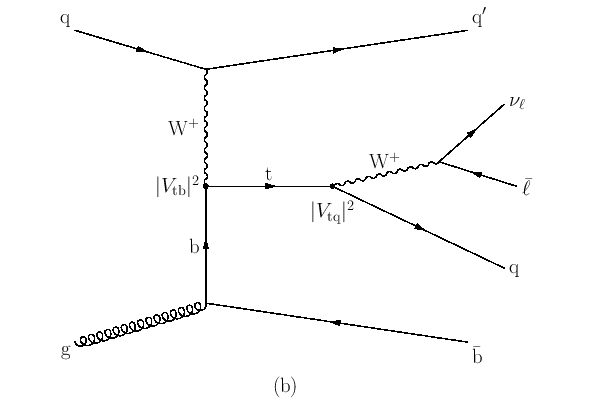} 
    \end{subfigure}
    \begin{subfigure}{0.4\textwidth}
        \includegraphics[width=\textwidth]{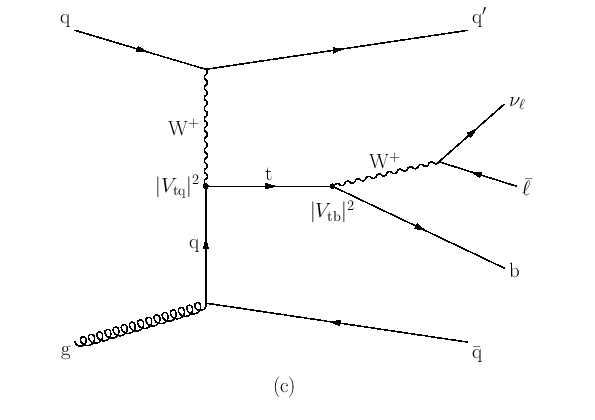} 
    \end{subfigure}
    \begin{subfigure}{0.4\textwidth}
        \includegraphics[width=\textwidth]{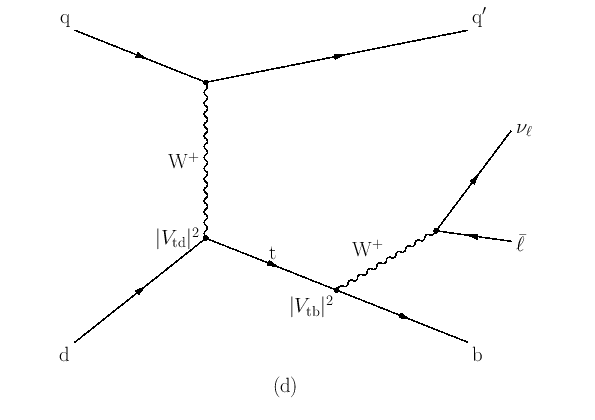} 
    \end{subfigure}
    \vspace{-0.3cm}
    \caption{Feynman diagrams for single top quark production via the $t-$channel with different contributions from the production and decay vertexes, that can be proportional to the CKM matrix elements $|V_{tb}|$, $|V_{ts}|$ or $|V_{td}|$~\cite{CMS_2}. 
    }
    \label{fig:CMS_2}
\end{figure}

In a different CMS analysis, the objective is to measure the CKM matrix elements $|V_{tb}|$, $|V_{ts}|$ and $|V_{td}|$ through the $t-$channel cross-section at $\sqrt{s} = 13$ TeV with data corresponding to an integrated luminosity of $L = 35.9$ fb$^{-1}$~\cite{CMS_2}. Events are selected if they contain one isolated electron or muon and have $m_T(W)>50$ GeV. Various signal topologies are considered, based on which vertex the Feynman diagram has in the single top production point and decay point. The vertexes can be either proportional to the $|V_{tb}|$ element or to the  $|V_{tq}|$ one\footnotemark[2], as shown in Figure \ref{fig:CMS_2}.            \footnotetext[1]{The normalised differential cross section as a function of $\cos\theta_{pol}$ at the parton level is related to the top quark polarisation $P$ as $\frac{1}{\sigma} \frac{d\sigma}{d\cos\theta_{pol}}=\frac{1}{2}(1+2A_l \cos(\theta_{pol}))$, where the polarising angle $\cos \theta_{pol}=\frac{\vec{p^*_l}\cdot \vec{p^*_q}}{|\vec{p^*_l}| |\vec{p^*_q}|}$ is computed in the top quark rest frame.} 
\footnotetext[2]{In this analysis only the quark referred to as $q$ indicates a light jet with $q \in \{u , d, s \}$ }

Different signal regions are selected based on the number of jets and $b-$tagged jets in the final state: two jets of which one $b-$tagged is the region enriched in events involving $|V_{tb}|$ in both production and decay, three jets of which one $b-$tagged is enriched in events involving $|V_{tb}|$ in only production or decay, and three jets of which two $b-$tagged is enriched in events with $|V_{tb}|$ in both production and decay. For each region, a BDT is trained and a maximum Likelihood fit is performed on their distributions. Confidence intervals are measured at 95\% CL in the SM scenario, assuming the SM unitarity constrain $|V_{tb}|+|V_{ts}|+|V_{td}|=1$, and they are found to be $|V_{tb}|>0.970$ and $|V_{td}|+|V_{ts}|< 0.057$. Two unconstrained Beyond the SM (BSM) scenarios are then considered, in the second one the assumption is that the mixing of the three families of quarks is negligible, and the measured values are $|V_{tb}|= 0.988 \pm 0.024$ and $|V_{td}|^2+|V_{ts}|^2=0.06 \pm 0.06$. 

\begin{figure}[htpb]
    \centering
    \includegraphics[width=0.4\textwidth]{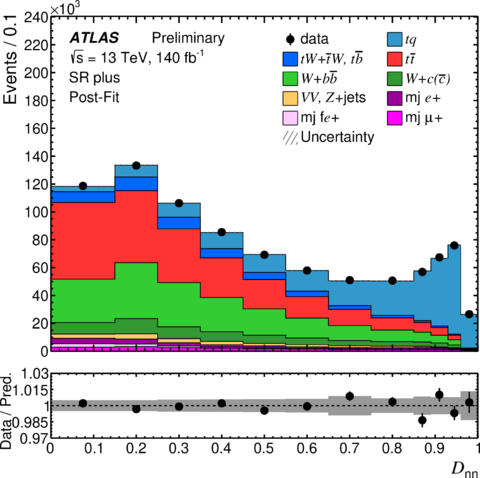}
    \includegraphics[width=0.4\textwidth]{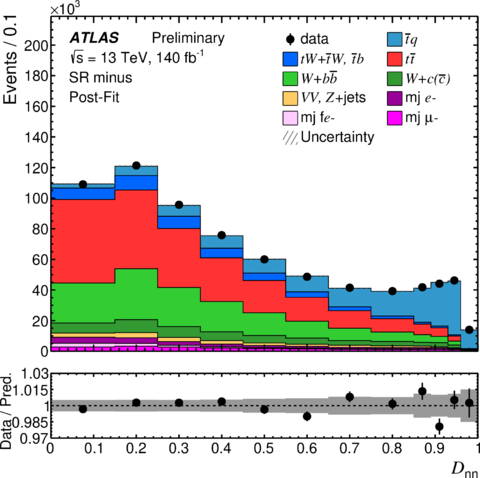}
    \vspace{-0.3cm}
    \caption{Post-fit NN output distribution in the $l^+$+jets and $l^-$+jets regions used for the single top $t-$channel cross-section measurement by the ATLAS Collaboration at $\sqrt{s}=13$~TeV~\cite{ATLAS_1}. 
    }
    \label{fig:ATLAS_1_NN}
\end{figure}

The ATLAS Collaboration has measured the inclusive single top $t-$channel cross-section at the centre of mass energy of $\sqrt{s} = 13$ TeV with data corresponding to an integrated luminosity of $L = 140$ fb$^{-1}$~\cite{ATLAS_1}. Events are selected if they contain one isolated electron or muon, two jets, one of which $b-$tagged. Further selection on the missing transverse energy $E_T^{miss}$, transverse momentum of the lepton $p_T(l)$, and $m_T(W)$ are applied to reduce multi-jet background and a selection on the invariant mass of the lepton and $b-$tagged jet $m(lb)$ is used to avoid bad modelling of top decays. To better separate signal versus background, a neural network (NN) is trained using 17 input kinematics variables of reconstructed objects, $W$ boson, and top quark. The NN is applied to data in two different regions based on the lepton charge and its distributions are used for the Profile likelihood fit to extract the total $t-$channel cross-section $\sigma_{t-chann}$, the top quark $\sigma_t$ and top anti-quark $\sigma_{\bar{t}}$ cross-sections and their ratio $R_t$, the expected and measured values are shown in Table~\ref{tab:ATLAS_1}. The NN distributions after the fit are shown in Figure~\ref{fig:ATLAS_1_NN}.

\begin{table}[htbp]
\begin{center}
    \begin{tabular}{l|cccc}  
     & $\sigma_t $[pb]       & $\sigma_{\bar{t}}$ [pb]   & $ \sigma_{t-chan}  $ [pb]  & $R_t = \sigma_t / \sigma_{\bar{t}}$ \\
    \hline
    Measured & $137 \pm 8$ & $ 84^{+6}_{-5}$ & $ 221 \pm 13$  & $1.636^{+0.036}_{-0.034}$    \\
    Predicted & $134.2 \pm 2.2$  &   $80.0 \pm 1.6$  &   $214.2 \pm 3.4$  & $1.677^{+0.010}_{-0.014}$\\
    \hline
    \end{tabular}
\caption{Expected and measured values of the single top $t-$channel cross-section measurement obtained by the ATLAS Collaboration at $\sqrt{s}=13$~TeV from a Profile Likelihood fit to the NN output distributions~\cite{ATLAS_1}.}
\label{tab:ATLAS_1}
\end{center}
\end{table}
The systematic uncertainties which have the highest impact on the cross-section measurements in this analysis are those relative to the top-quark production process modelling, such as the NLO-matching scale, the parton shower and final state radiation, and systematics relative to the jet energy scale and the tagging of jets originating from bottom quarks. In the ratio $R_t$ measurement many systematic uncertainties cancel out, and those that still have a big impact are those related to the $W+c$-jets background cross-section, the top quark parton shower and PDFs. 

\begin{figure}[htpb]
    \centering
    \begin{minipage}{0.4\textwidth}  
        \centering
        \includegraphics[width=0.5\textwidth]{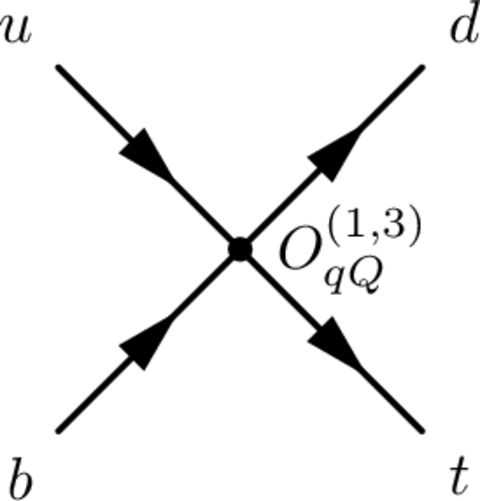}
            \caption{Feynman diagram of a four-quark contact interaction producing a single top quark, that is considered in the context of an EFT interpretation of the ATLAS Collaboration analysis in reference~\cite{ATLAS_1}.}
    \label{fig:ATLAS_1_diagram}
    \end{minipage}
    \hfill
    \begin{minipage}{0.55\textwidth}  
    \vspace{-0.2cm}
    Furthermore, the analysis extracts the value for the CKM matrix element $|V_{tb}|$ assuming that $|V_{tb}| \gg |V_{ts}|,|V_{td}|$ and the top quark always decay to a $W$ boson and a bottom quark, obtaining $f_{LV} \cdot V_{tb} = 1.016 \pm 0.031$. In a more generalised CKM interpretation the assumption on the contribution of $ |V_{ts}|,|V_{td}|$ is removed, and nine different contributions to the single top production are considered, based on the quark $q \in {d,s,b}$ that is present in the vertex $Wtq$ in production and decay. The setup used is to set one CKM element to 0 and perform a 2D maximum likelihood scan over the other two CKM elements. 
    \end{minipage}
\end{figure}

In the contest of Effective Field Theory (EFT) interpretation, dimension 6 operators are considered in this analysis to parameterise new physics effects, in particular the four-quark operator $O_{q,Q}^{(1,3)}$ which leads to single top quark production via non-SM interactions, as shown in Figure~\ref{fig:ATLAS_1_diagram}. A maximum Likelihood scan on the corresponding Willson coefficient is used to extract the 95\% CL interval that is determined to be: $-0.25 < C^{(1,3)}_{qQ} < 0.12$. 

\begin{figure}[htpb]
    \centering
    \includegraphics[width=0.4\textwidth]{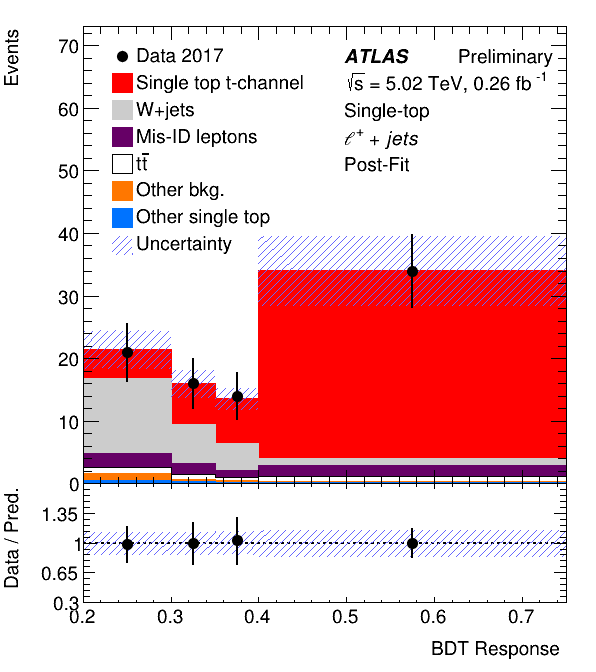}
    \includegraphics[width=0.4\textwidth]{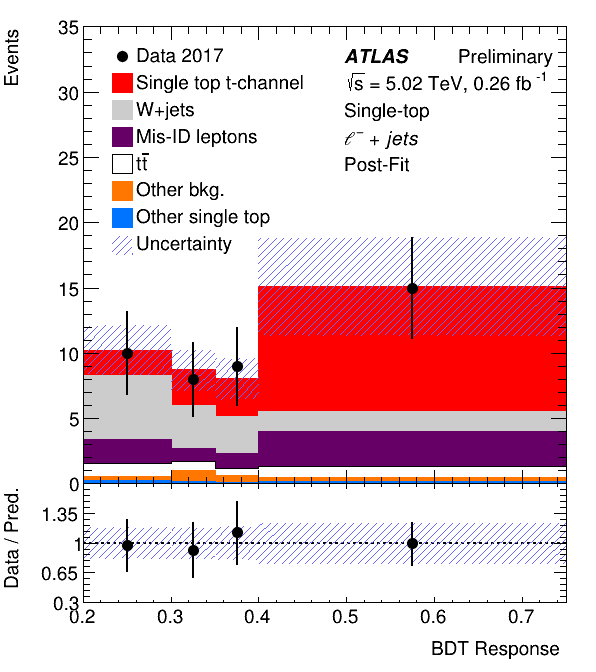}
    \vspace{-0.3cm}
    \caption{Post-fit BDT output distribution in the $l^+$+jets and $l^{-}$+jets regions used for the single top $t-$channel cross-section measurement by the ATLAS Collaboration at $\sqrt{s}=5.02$~TeV~\cite{ATLAS_2}. 
    }
    \label{fig:ATLAS_2}
\end{figure}

\newpage
The ATLAS Collaboration measured the $t-$channel single top production cross-section also at the centre-of-mass energy of $\sqrt{s} = 5.02$~TeV with data corresponding to an integrated luminosity of $257$~pb$^{-1}$~\cite{ATLAS_2}. The sample of data used has a low pile-up, with an average number of interactions per bunch crossing $<\mu>=2$. In this analysis, events are selected if they contain one isolated electron or muon, exactly two jets, one of which $b-$tagged, with a separation in pseudo-rapidity between the two jets $\Delta\eta(b,j)>1.5$. Further selections are applied on kinematic observables to reduce the mis-identified background and increase the signal purity. A BDT is trained on 9 input variables and a three-fold cross-validation is used to test the performances and check for over-training effects. A profile maximum likelihood fit is performed on the BDT output distribution in two different regions, defined based on the lepton charge, to extract the total $t-$channel cross-section $\sigma_{t-chann}$, the top quark $\sigma_t$ and top anti-quark $\sigma_{\bar{t}}$ cross-sections and their ratio $R_t$, the expected and measured values are shown in Table~\ref{tab:ATLAS_2}. 
\begin{table}[!h!tbp]
\begin{center}
    \begin{tabular}{l|cccc}  
     & $\sigma_t $[pb]       & $\sigma_{\bar{t}}$ [pb]   & $ \sigma_{t-chan}  $ [pb]  & $R_t = \sigma_t / \sigma_{\bar{t}}$ \\
    \hline
    Measured & $19.5^{+4.8}_{-3.8}$ & $  7.1^{+4.3}_{-2.6}$ & $ 26.6^{+6.1}_{-5.4}$  & $ 2.74^{+1.78}_{-0.88}$    \\
    Predicted &  $20.3^{+0.5}_{-0.4}$  &   $10.0^{+0.2}_{-0.3}$  &  $30.3^{+0.5}_{-0.5}$  &  $2.03^{+0.06}_{-0.07}$\\
    \hline
    \end{tabular}
\caption{Expected and measured values of the single top $t-$channel cross-section measurement obtained by the ATLAS Collaboration at $\sqrt{s}=5.02$~TeV from a Profile Likelihood fit to the BDT output distributions~\cite{ATLAS_2}.}
\label{tab:ATLAS_2}
\end{center}
\end{table}
Figure~\ref{fig:ATLAS_2} shows the BDT output distributions after the fit. The single top $t-$channel production process is observed with a significance of $6.1$ standard deviations at $\sqrt{s}=5.02$~TeV with respect to the background-only hypothesis. The total uncertainty on this cross-section measurement has a similar contribution from the statistical and systematic error, while the ratio measurement is statistically dominated. 

\section{Single top $s-$channel production}
Measurements of single top production in the $s-$channel are very challenging at the LHC because of its little production cross-section ( $\sigma_{s-channel}(13 \text{TeV}) = 10.32^{+0.40}_{-0.32}$~pb) and because it is challenging to separate it from its main background (top quark pair production). The ATLAS Collaboration has observed evidence of this process using the full Run 2 dataset at an energy in the centre-of-mass of $\sqrt{s}=13$~TeV~\cite{ATLAS_3}. Events are selected if they contain one electron or muon, $E_T^{miss}>35$~GeV and $m_T(W)>30$~GeV and they contain at least two $b-$tagged jets. The Matrix Element Method (MEM)~\cite{MEM} is used to separate signal events from background ones: it is a per-event likelihood calculation that returns the probability that a certain final state $X$ is of the process $H_{proc}$ ($P(X|H_{proc})$). 

\begin{figure}[htpb]
    \begin{minipage}{0.51\textwidth}
    \centering
        \includegraphics[width=0.85\textwidth]{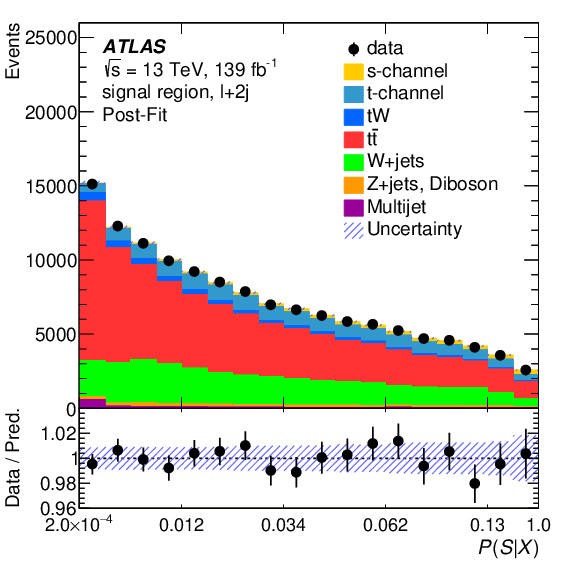}
        \vspace{-0.2cm}
        \caption{Distribution of the MEM discriminant used by the ATLAS Collaboration for the $s-$channel cross-section measurement~\cite{ATLAS_3}. 
        }
        \label{fig:ATLAS_3}
    \end{minipage}
    \hfill
    \begin{minipage}{0.43\textwidth}
        \vspace{-1.5cm}
         Then applying Bayes' theorem a discriminant is obtained: its distribution is shown in Figure~\ref{fig:ATLAS_3}. The measured cross-section is $\sigma=8.2^{+3.5}_{-2.9}$~pb, and the background-only hypothesis is rejected with an observed significance of $3.3$ standard deviations. In the cross-section measurement the systematic uncertainties with the highest impact are related to the $t\bar{t}$ background normalisation and modelling, the $s-$channel modelling, jet energy resolution and scale and MC statistics.
    \end{minipage}
\end{figure}

\newpage

\section{Single top $W$ associated production}
\begin{figure}[!htpb]
    \centering
    \includegraphics[width=0.45\textwidth]{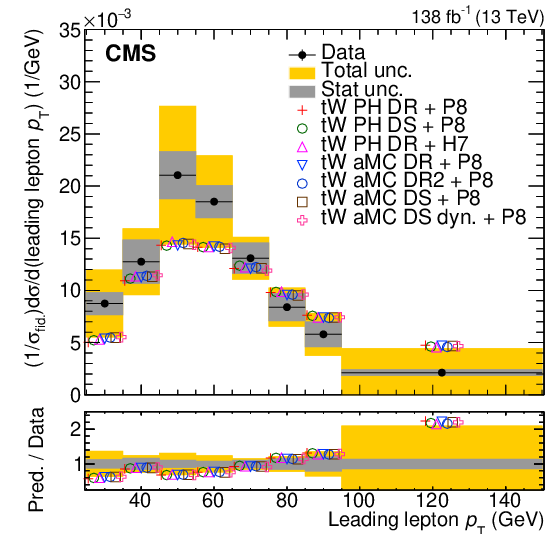}
    \caption{Normalised fiducial differential $tW$ production cross section as a function of the $p_T$ of the leading lepton measured by the CMS Collaboration at $\sqrt{s}=13$~TeV~\cite{CMS_3}.}
    \label{fig:CMS_3}
\end{figure}

In this section two recent CMS measurements of the single top $W$ associated production ($tW$) are reported: one is in the dileptonic decay channel, while the other is in the single-lepton decay channel. The first one is the differential $tW$ production cross-section measurement at $\sqrt{s}=13$~TeV with data corresponding to an integrated luminosity of $L=138$~fb$^{-1}$~\cite{CMS_3}. Events are selected if they have one electron and one muon of opposite charge in the final state, and their invariant mass is greater than $m(e\mu) > 20$~GeV. For the inclusive cross-section measurement, selected events are separated into three orthogonal samples: one with one $b-$tagged jet, one with two jets one of which $b-$tagged, and a control region for the $t\bar{t}$ background. A profile likelihood fit is performed on the output distribution of a BDT applied to the first two samples defined and to the sub-leading jet transverse momentum in the control region. The measured $tW$ cross-section is $\sigma_{tW}=79.2 ^{+7.8}_{-8.1}$~pb. Differently in the case of the differential measurements, only the signal region with events that contain one $b-$tagged jet is considered with the further request to have no other jets with $20$~GeV~$< p_T < 30$~GeV. A profile likelihood unfolding is used to obtain normalised particle-level differential cross-sections in a dedicated fiducial volume. Differential distributions are found to be in good agreement with Monte Carlo predictions, even if a slight disagreement is found for the $\Delta \Phi (e,\mu)$ distribution and for the leading lepton $p_T$ as shown in Figure~\ref{fig:CMS_3}.

\begin{figure}[htpb]
    \centering
    \includegraphics[width=0.4\textwidth]{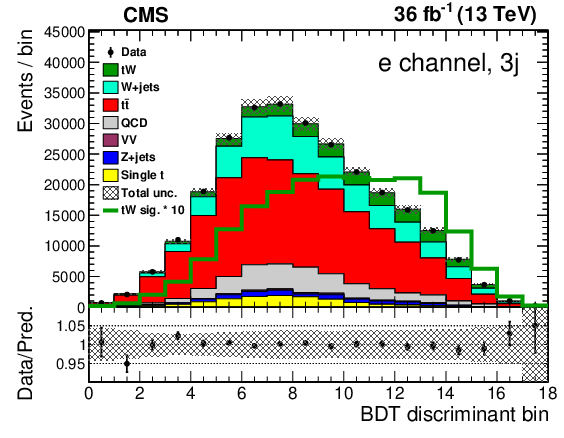}
    \includegraphics[width=0.4\textwidth]{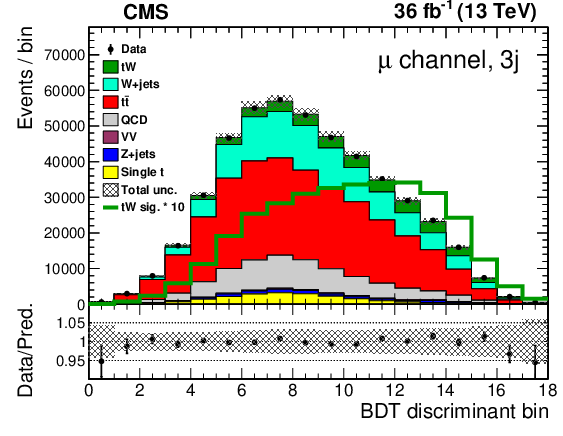}
    \caption{Post-fit BDT output distribution in the electron and muon channels used for the $W$ associated single top cross-section measurement in the single lepton decay channel by the CMS Collaboration at $\sqrt{s}=13$~TeV~\cite{CMS_4}. 
    }
    \label{fig:CMS_4}
\end{figure}

A different analysis has observed the $tW$ associated production in the single-lepton decay channel at an energy in the centre-of-mass of $\sqrt{s}=13$~TeV with data corresponding to an integrated luminosity of $L=36$~fb$^{-1}$~\cite{CMS_4}. Events are selected if they have one electron or muon and at least two jets, one of which must be $b-$tagged. Events are divided based on the number of jets in the final state: three jets for the signal region, two jets for events in the control region for the $W+jets$ background and 4 jets for the $t\bar{t}$ production background control region. The $W+jets$ background normalisation and the multi-jet background normalisation and template are extracted from the data. Two BDTs are trained on events with electrons or muons in the final state to separate the $tW$ signal from the $t\bar{t}$ background production and their output distributions are used for a maximum likelihood fit, as shown in Figure~\ref{fig:CMS_4}.

The measured cross-section is $\sigma_{tW} = 89 \pm 13$~pb which is compatible with the SM prediction.

\section{Conclusions}
In this report, an overview of recent single top results by the ATLAS and CMS Collaborations using $pp$ collisions collected during the Run 2 of the LHC has been given. Considering only the inclusive measurements included in this work, the relative uncertainties reached for the three main production modes of single top are: a $5.9\%$ relative uncertainty for the $t-$channel cross-section and a $2.2\%$ for the top quark and anti-quark ratio in~\cite{ATLAS_1}, a relative uncertainty of $+42\%$ $-35\%$ on the $s-$channel cross-section in~\cite{ATLAS_3}, and a relative uncertainty of $10\%$ on the $tW$ production in~\cite{CMS_3}. 
In the analysis discussed in this report, we can see that the leading systematic uncertainties are: signal and background modelling and jet-related uncertainties in the $t-$channel and $s-$channel single top measurements and background normalization and jet energy scale for the $tW$ single top production.

These results show that we have gained a deep understanding of the single top production processes and reached good precision, especially in the $t-$channel single top measurements. In general, all measurements are found to be in agreement with the SM predictions both for inclusive and differential results.

\newpage

\bibliography{eprint}{}
\bibliographystyle{unsrt}
 
\end{document}